\input{aipcheck}

\documentclass[
    ,final            % use final for the camera ready runs
  ]
  {aipproc}

\usepackage{amsmath,epstopdf}
\usepackage{amsfonts,amssymb,amsthm}

\layoutstyle{8x11single}

\begin{document}

\title{On the Harmonic Scattering Amplitudes from a Nonlinear Elastic Spherical Inclusion}


\author{Christopher M. Kube}{
  address={Bennett Aerospace, Inc., 1249 Kildaire Farm Rd., Cary, NC 27511, USA}
}

\author{Andrea P. Arguelles}{
  address={Mechanical and Materials Engineering, University of Nebraska-Lincoln, Lincoln, NE, 68588, USA}
}

\author{Brandon McWilliams}{
  address={Weapons and Materials Research Directorate, U.S. Army Research Laboratory, Aberdeen Proving Ground, MD, 21001, USA}
}

\begin{abstract}
In this paper, the interaction of an incident finite amplitude longitudinal wave with a localized region of nonlinearity is considered. This interaction produces a secondary field represented by a superposition of first-, second-, and third-harmonic components. The secondary field is solely a result of the quadratic and cubic elastic nonlinearity present within the region of the inclusion. The second-harmonic scattering amplitude depends on the quadratic nonlinearity parameter $\beta$, while the first- and third-harmonic amplitudes depend on the cubic nonlinearity parameter $\gamma$. The special cases of forward and backward scattering amplitudes were analyzed. For each harmonic, the forward-scattering amplitude is always greater than or equal to the backward scattering amplitude in which the equality is only realized in the Rayleigh scattering limit. Lastly, the amplitudes of the scattered harmonic waves are compared to predicted harmonic amplitudes derived from a plane wave model.
\end{abstract}

\maketitle

%%%%%%%%%%%%%%%%%%%%%%%%%%%%%%%%%%%%%%%%%%%%
%% MAINMATTER
%%%%%%%%%%%%%%%%%%%%%%%%%%%%%%%%%%%%%%%%%%%%

\section{Introduction}

Recently, a number of investigators have considered the interaction of an elastic wave with a localized region of elastic nonlinearity \cite{Tang12,Wang16,Kube17a}. Tang \textit{et al.} modeled the interaction of an incident longitudinal wave with a region with spatially-dependent quadratic nonlinearity, which was contained in a linear elastic host medium \cite{Tang12}. In Tang \textit{et al.}, the problem was formulated in a scattering framework in which the region of nonlinearity was denoted as an inclusion. Wang and Achenbach modeled the interaction of an incident torsional guided wave on a local annular region of cubic nonlinearity \cite{Wang16}. Solutions based on reciprocity found backward and forward propagating waves from the region of nonlinearity, which contained first- and third-harmonic components \cite{Wang16}. Kube then considered the interaction of an incident longitudinal or shear wave with a region of spatially-dependent and generally anisotropic quadratic and cubic nonlinearity encompassed in an otherwise linear host medium \cite{Kube17a}. The secondary field generated from the region of nonlinearity contained first-, second-, and third-harmonic components. The second-harmonic component was found to be a result of the quadratic nonlinearity while the first- and third-harmonic components were generated by the effects of cubic nonlinearity \cite{Kube17a}, which parallels the finding of Wang and Achenbach \cite{Wang16}.

\medskip

This paper considers the interaction of an incident longitudinal wave with a spherical region of constant quadratic and cubic elastic nonlinearity. Solutions to this problem arrive from the more general model of Ref. \cite{Kube17a}. This example describes the situation in which localized nonlinearity exhibits spherical symmetry, e.g., a point-like stress concentration. It is assumed that the macroscopic linear elastic properties are not affected by the microstructure evolution that led to the region of nonlinearity. Throughout, the region of nonlinearity is denoted as an inclusion while the generated secondary field is labeled the scattered field. The displacement amplitudes of the forward and backward scattered harmonics are given special attention because of their application to commonly used experimental configurations, i.e., co-linear \textit{through transmission} or normal incidence \textit{pulse-echo}. Lastly, the scattered displacement amplitudes are compared to plane wave displacement solutions to highlight their differences.

\section{Theory}

Here, we consider a special simplified case of the more general model of Ref. \cite{Kube17a}. A longitudinal wave with initial amplitude $U$, wave number $k$, and angular frequency $\omega$ is incident on a spherical inclusion of radius $a$. The inclusion is isotropic and described by a hyperelastic strain energy function that is a Taylor expansion of the Lagrangian strain tensor $E_{ij}$ up to quartic order, $W=\frac{1}{2!}C_{ijkl}E_{ij}E_{kl}+\frac{1}{3!}C_{ijklmn}E_{ij}E_{kl}E_{mn}+\frac{1}{4!}C_{ijklmnpq}E_{ij}E_{kl}E_{mn}E_{pq}$. The medium surrounding the inclusion is also isotropic with the same linearly elastic modulus $C_{ijkl}$. The scattered displacement field generated by the interaction of the incident longitudinal wave and the inclusion is \cite{Kube17a}
\begin{align}
u_i^S&=\hat{r}_{i}\frac{\gamma U^3k}{8 r\left(2-2\cos\theta\right)^{3/2}}\left[\sin\left(ka\sqrt{2-2\cos\theta}\right)-ka\sqrt{2-2\cos\theta}\cos\left(ka\sqrt{2-2\cos\theta}\right)\right]\cos\left(kr-\omega t\right)\notag\\
&+\hat{r}_{i}\frac{\beta U^2}{16r\left(2-2\cos\theta\right)^{3/2}} \left[\sin\left(2ka\sqrt{2-2\cos\theta}\right)-2ka\sqrt{2-2\cos\theta}\cos\left(2ka\sqrt{2-2\cos\theta}\right)\right]\sin\left(2kr-2\omega t\right)\notag\\
&-\hat{r}_{i}\frac{\gamma U^3k}{216 r\left(2-2\cos\theta\right)^{3/2}}\left[\sin\left(3ka\sqrt{2-2\cos\theta}\right)-3ka\sqrt{2-2\cos\theta}\cos\left(3ka\sqrt{2-2\cos\theta}\right)\right]\cos\left(3kr-3\omega t\right)\label{Eq1},
\end{align}
where $r$ is the distance from the center of the inclusion and the observation point, $\theta$ is the angle formed between the incident and scattered wave directions. Additional details on the derivation of Equation (\ref{Eq1}) can be located in Ref. \cite{Kube17a}. Equation (\ref{Eq1}) permits the determination of scattering amplitudes for any direction. In the present paper, we will focus on backward and forward scattering directions given by the conditions $\theta=\pi$ and $\theta=0$, respectively. The quadratic and cubic nonlinearity parameters found in Equation (\ref{Eq1}) are
\begin{flalign}
\hspace{3em}&\beta=\frac{1}{c_{11}}\left(3c_{11}+c_{111}\right),\text{ and}\label{Eq2}\\
\hspace{3em}&\gamma=\frac{1}{c_{11}}\left(3c_{11}+6c_{111}+c_{1111}\right),\label{Eq3}&
\end{flalign}
respectively. Here, $c_{11}$, $c_{111}$, and $c_{1111}$ are the second-, third-, and fourth-order elastic stiffnesses, respectively. These elastic stiffnesses can be converted to the convention of Landau and Lifshitz \cite{Landau86} via the relations $c_{11}=\lambda+2\mu$, $c_{111}=2\mathcal{A}+6\mathcal{B}+2\mathcal{C}$, and $c_{1111}=24\left(\mathcal{E}+\mathcal{F}+\mathcal{G}+\mathcal{H}\right)$. The fourth-order stiffnesses $\mathcal{E}$, $\mathcal{F}$, $\mathcal{G}$, and $\mathcal{H}$ were introduced by Zabolotskaya \cite{Zabolotskaya86}. The definition of $\beta$, derived in the scattering formalism, is equivalent to the traditional nonlinearity parameter associated with plane wave propagation. The cubic nonlinearity parameter $\gamma$ defined here differs by a factor of 2 from the cubic nonlinearity associated with plane wave propagation \cite{Kube17b}.

\medskip

\begin{figure}[!t]
  \includegraphics[scale=1]{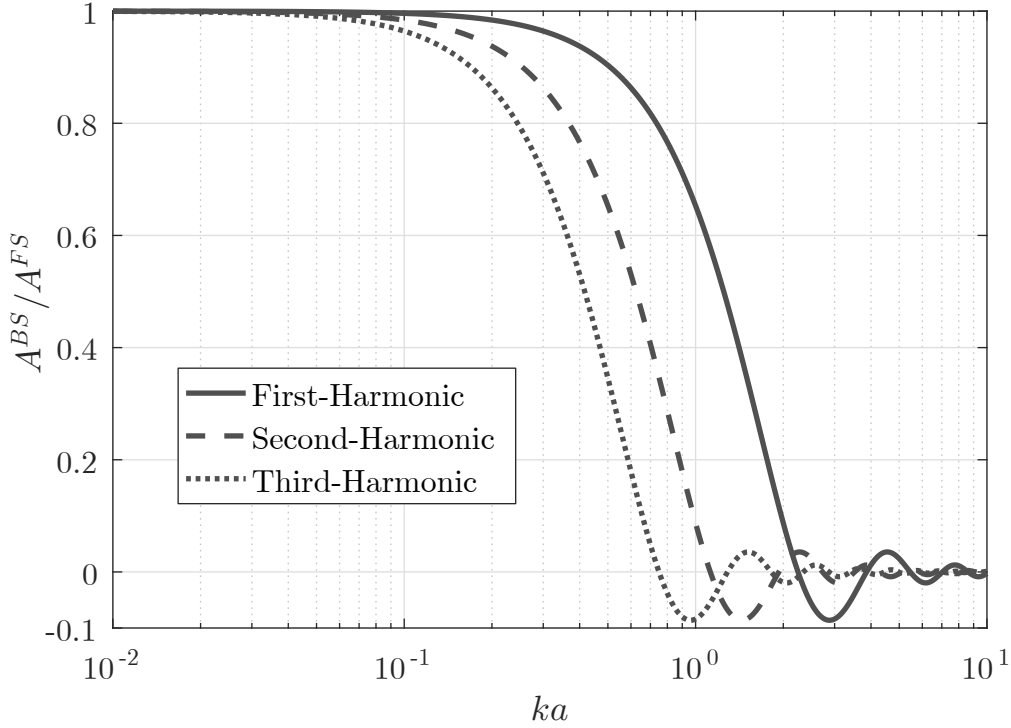}
  \caption{Comparison of the forward and backward scattering amplitudes for each of the harmonics as a function of $ka$.}
  \label{Fig1}
\end{figure}

Bulk wave ultrasonic harmonic generation experiments historically involve a colinear through transmission configuration, which exploits the cumulative behavior of self-resonance (see Ref. \cite{Breazeale84} for an example set-up). More recently, researchers have attempted to recover the quadratic nonlinearity parameter by measurement of the second-harmonic in a pulse-echo configuration \cite{Best14,Jeong16}. These measurements attempt to measure the second-harmonic from a reflected back surface echo. This method relies on the non-plane wave behavior of waves generated from real ultrasonic transducers; a true plane wave phase inverts upon reflection at a pressure-release boundary, which causes the second-harmonic component to vanish at the source. The familiarity of pulse-echo and through transmission configurations for nonlinearity measurements indicates that the forward and backward scattering amplitudes will be of practical interest. The harmonic amplitudes of the forward scattered field are easily found from Equation. (\ref{Eq1}) by taking the limit as $\theta\rightarrow 0$, which gives
\begin{flalign}
&\hspace{3em}A_1^{FS}=A_3^{FS}=\gamma\frac{U^3k^4a^3}{24r},\label{Eq4}\\
&\hspace{3em}A_2^{FS}=\beta\frac{U^2k^3a^3}{6r}.\label{Eq5}&
\end{flalign}
The forward scattering amplitudes of the first- and third-harmonic are equal. All three forward scattering amplitudes are proportional to the volume of the inclusion. An asymptotic equality of the first- and third-harmonics is also observed in the Rayleigh scattering limit, $ka\ll1$. The corresponding backward scattering amplitudes are found by letting $\theta=\pi$ in Equation (\ref{Eq1}), which are
\begin{flalign}
&\hspace{3em}A_1^{BS}=\gamma\frac{U^3 k}{64r}\left[\sin\left(2ka\right)-2ka\cos\left(2ka\right)\right],\label{Eq6}\\
&\hspace{3em}A_2^{BS}=\beta\frac{U^2}{128r}\left[\sin\left(4ka\right)-4ka\cos\left(4ka\right)\right],\text{ and}\label{Eq7}\\
&\hspace{3em}A_3^{BS}=\gamma\frac{U^3 k}{1728r}\left[\sin\left(6ka\right)-6ka\cos\left(6ka\right)\right].\label{Eq8}&
\end{flalign}
The inclusion radius $a$ causes the backward scattering amplitude to exhibit oscillating growth as $ka$ increases, whereas the forward scattering amplitude has a power-law dependence on $a$ and $k$. Ratios of the backward and forward scattering amplitudes were formed in order to observe their relative sensitivity. The $ka$ dependence of the ratios are illustrated in Figure \ref{Fig1}.
In the Rayleigh scattering limit, $ka\ll1$, the forward and backward scattering amplitudes are equal, which indicates the well-known isotropic scattering behavior at small $ka$. The scattering behavior increasingly favors the forward direction as $ka$ moves further from the Rayleigh limit. Thus, the forward scattered harmonics will tend to be easier resolved than the backward scattered harmonics. In the following section, a hypothetical experimental configuration is analyzed in order to further the comparison between forward and backward scattering.

\medskip

In some instances, the spatial distribution or extent of a nonlinear region, e.g. a region of fatigue damage, is uncertain. A plain wave analysis is most appropriate when the extent of region of nonlinearity is uniform and extends throughout the structure, or at least within the volume containing the wave. Conversely, the scattering theory is appropriate when it is safe to assume that the nonlinearity is localized in space. To assess the potential error associated with spatial uncertainty of the nonlinearity, the following section provides a practical comparison between the forward and backward scattering amplitudes and displacement amplitudes of harmonics derived through a plane wave theory. The harmonic amplitudes from the plane wave theory are observed from the general displacement solution \cite{Kube17b},
\begin{flalign}
\hspace{3em}u_i^{PW}&=\frac{1}{32}\hat{u}_iU\sqrt{\left[32+\left(\beta Uk^2x\right)^2\right]^2+4k^6U^4x^2\left(\beta^{^2}+\gamma \right)^2}\sin\left(kx-\omega t\right)\notag\\
&\quad+\frac{1}{8}\hat{u}_i\beta U^2k^2x\cos\left[2\left(kx-\omega t\right)\right]+\frac{1}{32}\hat{u}_i\beta^2U^3k^4x^2\sqrt{\kappa}\sin\left[3\left(kx-\omega t\right)\right],\label{Eq9}&
\end{flalign}
where $x$ is the propagation distance and $\kappa=1+16/\left(3kx\right)^2\left[1+\gamma/\left(2\beta^2\right)+\gamma^2/\left(4\beta^4\right)\right]$. The three harmonic amplitudes are
\begin{flalign}
&\hspace{3em}A_1^{PW}=U\left(1+\frac{\beta^2U^2k^4x^2}{32}\right),\label{Eq10}\\
&\hspace{3em}A_2^{PW}=\beta\frac{U^2k^2x}{8},\label{Eq11}\\
&\hspace{3em}A_3^{PW}\approx\gamma\frac{U^3k^3x}{48}\text{ for }\kappa\approx 16/\left(3kx\right)^2,\text{ and}\label{Eq12}\\
&\hspace{3em}A_3^{PW}\approx \beta^2\frac{U^3k^4x^2}{32}\text{ for }\kappa\approx1\label{Eq13}&
\end{flalign}
where, for the isotropic symmetry considered, $\beta$ and $\gamma$ are defined in Equation (\ref{Eq2}) and (\ref{Eq3}), respectively. Note that $\gamma$ in Equation (\ref{Eq3}) differs by a factor 2 from its plane wave definition given in Ref. \cite{Kube17b}; the associated expressions containing $\gamma$ [Equation (\ref{Eq9}), (\ref{Eq12}), and the definition of $\kappa$] were appropriately scaled to account for this difference. In arriving at Equation (\ref{Eq10}), it was assumed that $32+\left(\beta Uk^2x\right)^2\gg2k^3U^2x\left(\beta^{^2}+\gamma \right)$, which is thought to always hold in practical experimental configurations.

\section{Example}

As a practical example, consider the following experimental configuration: $U=1$ nm, $f=10$ MHz, $v=5.9$ mm/$\mu$s, $a=2$ mm, $r=10$ mm, $\beta=30$, $\gamma=150$, which represents a plausible experiment conducted on a steel sample. The following forward and backward scattering amplitudes were calculated from Equation (\ref{Eq4}-\ref{Eq8}): $A_1^{FS}=A_3^{FS}=6.43\times10^{-8}$ nm, $A_2^{FS}=4.83\times10^{-3}$ nm, $A_1^{BS}=2.22\times10^{-11}$ nm, $A_2^{BS}=1.85\times10^{-6}$ nm, and $A_3^{BS}=6.34\times10^{-12}$ nm. Now, instead of using the scattering formulas, the plane wave formulas are applied with a propagation distance of $x=2a=4$ mm. This represents the situation of cumulative harmonic growth over a 4 mm propagation distance. The following plane wave amplitudes were calculated: $A_1^{PW}=1.0000058$ nm, $A_2^{PW}=1.70\times10^{-3}$ nm, and $A_3^{PW}=5.79\times10^{-6}$ nm. $A_3^{PW}$ was calculated from Equation (\ref{Eq13}) with $\kappa=1.0011$. Several items are noted,
\begin{itemize}
\item The forward scattered second-harmonic amplitude was found to be 2.84 times greater than the plane wave second-harmonic amplitude. This result gives an example of the potential error present when $\beta$ is calculated from the plane wave formula when it is known that the material nonlinearity is localized.
\item The forward scattered first-harmonic and third-harmonic amplitudes ($A_1^{FS}=A_3^{FS}$) are very small, approximately 8 orders of magnitude smaller than the initial fundamental primary wave amplitude $U$. Also, $A_1^{FS}$ and $A_3^{FS}$ are about 2 orders of magnitude smaller than $A_3^{PW}$, which is another indicator of potential error when applying the plane wave formula.
\item The backward scattered harmonics are all very small. The second-harmonic amplitude ($A_2^{BS}=1.85\times10^{-6}$ nm) is thought to be the only backward scattered harmonic with measurement plausibility (with optimized experimental parameters).
\end{itemize}

\section{Conclusion}

In this paper, the interaction of an incident finite amplitude longitudinal wave with a localized spherical region of elastic nonlinearity was considered. The secondary field produced by the interaction resembles a scattered field, which contains a superposition of first-, second-, and third-harmonic components. The expression for the scattered field given in Equation (\ref{Eq1}) accounts for scattering into any direction in space. Special consideration was made for the forward and backward scattering amplitudes because of potential measurements using a through transmission or pulse-echo configuration. It was found that the forward scattered amplitudes is always greater than or equal to the backward scattering amplitude. The backward scattering amplitude is very weak except in the Rayleigh limit in which $ka\ll1$, which is especially prohibitive for usual experimental configurations. A hypothetical experimental example was provided in order to evaluate and compare forward and backward scattering amplitudes. For this example, only the forward-scattered second-harmonic component displayed measurement plausibility. Lastly, this example provided a quantitative analysis of potential error when assuming that the measured harmonic components can be described using a plane wave model.

\section{Acknowledgments}

The research reported in this document was performed under contract(s)/instrument(s) W911QX-16-D-0014 with the U.S. Army Research Laboratory. The views and conclusions contained in this document are those of Bennett Aerospace and the U.S. Army Research Laboratory. Citation of manufacturer's or trade names does not constitute an official endorsement or approval of the use thereof. The U.S. Government is authorized to reproduce and distribute reprints for Government purposes notwithstanding any copyright notation hereon.

\bibliographystyle{aipproc}   % if natbib is available

\end{document}